\documentclass[longauth]{aa}  

\usepackage{graphicx}
\usepackage{txfonts}
\usepackage{color}
\definecolor{yaleblue}{rgb}{0.06, 0.3, 0.57}
\usepackage{hyperref}
\hypersetup{colorlinks=true,
                linkcolor=black,
                urlcolor=blue,
                citecolor=yaleblue}
\setcitestyle{notesep={ }}
\begin{document} 

\title{{\it Euclid:} Superluminous supernovae in the Deep Survey$\thanks{This paper is published on behalf of the {\it Euclid} Consortium}$}

\author{Inserra C.\inst{1,2}, Nichol R. C.\inst{3}, Scovacricchi D.\inst{3}, Amiaux J.\inst{4}, Brescia M.\inst{5}, Burigana C.\inst{6,7,8}, Cappellaro E.\inst{9}, Carvalho C. S. \inst{30}, Cavuoti S.\inst{5,11,12}, Conforti V.\inst{13}, Cuillandre J.-C. \inst{4,14,15}, da Silva A.\inst{10,16}, De Rosa A.\inst{13}, Della Valle M.\inst{5,17}, Dinis J.\inst{10,16},  Franceschi E.\inst{13}, Hook I.\inst{18}, Hudelot P.\inst{19}, Jahnke K.\inst{20}, Kitching T.\inst{21}, Kurki-Suonio H.\inst{22}, Lloro I.\inst{23}, Longo G.\inst{11,12}, Maiorano E. \inst{13}, Maris M. \inst{24}, Rhodes J. D.\inst{25}, Scaramella R.\inst{26}, Smartt S. J.\inst{2}, Sullivan M.\inst{1}, Tao C.\inst{27,28},  Toledo-Moreo R.\inst{29}, Tereno I. \inst{16,30}, Trifoglio M.\inst{13} and Valenziano L.\inst{13}}

   \institute{Department of Physics and Astronomy,  University of Southampton, Southampton, SO17 1BJ\\
              \email{C.Inserra@soton.ac.uk}
              \and
              Astrophysics Research Centre, School of Mathematics and Physics, Queen's University Belfast, Belfast BT7 1NN, UK
              \and
             Institute of Cosmology and Gravitation, University of Portsmouth, Portsmouth PO1 3FX, UK
             \and
             IRFU, CEA, Universit\'e Paris-Saclay F-91191 Gif-sur-Yvette Cedex, France
             \and
             INAF-Capodimonte Observatory, Salita Moiariello 16, I-80131, Napoli, Italy
             \and
             INAF, Istituto di Radioastronomia, Via Piero Gobetti 101, I-40129 Bologna, Italy
             \and
             Dipartimento di Fisica e Scienze della Terra, Universit\`a degli Studi di Ferrara, Via Giuseppe Saragat 1, I-44122 Ferrara, Italy
             \and
             INFN - Bologna, Via Irnerio 46, I-40126, Bologna, Italy
             \and
             INAF - Osservatorio Astronomico di Padova, Vicolo dell'Osservatorio 5, Padova I-35122, Italy
             \and
             Instituto de Astrof\'isica e Ci\^encias do Espa\c co, Faculdade de Ci\^encias, Universidade de Lisboa, Campo Grande, PT-1749-016 Lisboa, Portugal
             \and
             INFN section of Naples, via Cinthia 6, I-80126, Napoli, Italy
             \and
             Department of Physics "E. Pancini", University Federico II, via Cinthia 6, I-80126, Napoli, Italy
             \and
             INAF-IASF Bologna, Via Piero Gobetti 101, I-40129 Bologna, Italy
             \and
             Universit\'e Paris Diderot, AIM, Sorbonne Paris Cit\'e, CEA, CNRS F-91191 Gif-sur-Yvette Cedex, France
             \and
             Observatoire de Paris, PSL Research University 61, avenue de l'Observatoire, F-75014 Paris, France
             \and
             Departamento de F\'{\i}sica, Faculdade de Ci\^encias, Universidade de Lisboa, Edif\'{\i}cio C8, Campo Grande, PT1749-016 Lisboa, Portugal
             \and
             International Center for Relativistic Astrophysics, Piazzale della Repubblica 2, I-65122 Pescara, Italy
             \and
             Department of Physics, Lancaster University, Lancaster, United Kingdom, LA1 4YB
             \and
             Institut  d'Astrophysique  de  Paris,  98bis  Boulevard  Arago,  F-75014, PARIS, France
             \and
             Max-Planck-Institut f\"ur Astronomie, K\"onigstuhl 17, D-69117 Heidelberg, Germany
             \and
             Mullard Space Science Laboratory, University College London, Holmbury St Mary, Dorking, Surrey RH5 6NT, UK
             \and
             Department of Physics and Helsinki Institute of Physics, Gustaf H\"allstr\"omin katu 2, 00014 University of Helsinki, Finland
             \and
             Institute of Space Sciences (IEEC-CSIC), c/Can Magrans s/n, 08193 Cerdanyola del Vall\`es, Barcelona, Spain
             \and
             INAF - Osservatorio Astronomico di Trieste Via G. B. Tiepolo 11 34131 Trieste Italy
             \and
             NASA Jet Propulsion Laboratory, California Institute of Technology, 4800 Oak Grove Drive, MS 169-237, CA, 91109, USA
             \and
             INAF - Osservatorio Astronomico di Roma, via Frascati 33, I-00078 Monteporzio Catone, Italy
             \and 
             Aix-Marseille Univ, CNRS/IN2P3, CPPM, Marseille, France
             \and
             Tsinghua Center for Asttrophysics, Tsinghua University, Beijing, China
             \and
             Depto. de Electro\'onica y Tecnolog\'ia de Computadoras Universidad Polit\'ecnica de Cartagena, 30202, Cartagena, Spain
             \and
             Instituto de Astrof\'{\i}sica e Ci\^encias do Espa\c o, Universidade de Lisboa, Tapada da Ajuda, PT-1349-018 Lisboa, Portugal
             }

   \date{Received XXXX}

 
  \abstract
   {In the last decade, astronomers have found a new type of supernova called `superluminous supernovae' (SLSNe) due to their high peak luminosity and long light-curves. These hydrogen-free explosions (SLSNe-I) can be seen to $z\sim4$ and therefore, offer the possibility of probing the distant Universe.}
   {We aim to investigate the possibility of detecting SLSNe-I using ESA's {\it Euclid} satellite, scheduled for launch in 2020. In particular, we study the {\it Euclid} Deep Survey (EDS) which will provide a unique combination of area, depth and cadence over the mission.}
   {We estimated the redshift distribution of {\it Euclid} SLSNe-I using the latest information on their rates and spectral energy distribution, as well as known {\it Euclid} instrument and survey parameters, including the cadence and depth of the EDS. To estimate the uncertainties, we calculated their distribution with two different set-ups, namely optimistic and pessimistic, adopting different star formation densities and rates. We also applied a standardization method to the peak magnitudes to create a simulated Hubble diagram to explore possible cosmological constraints.}
   {We show that {\it Euclid} should detect approximately 140 high-quality SLSNe-I to $z \sim 3.5$ over the first five years of the mission (with an additional 70 if we lower our photometric classification criteria). This sample could revolutionize the study of SLSNe-I at $z>1$ and open up their use as probes of star-formation rates, 
galaxy populations, the interstellar and intergalactic medium. In addition, a sample of such SLSNe-I could improve constraints on a time-dependent dark energy equation-of-state, namely $w(a)$,  when combined with local SLSNe-I and the expected SN Ia sample from the Dark Energy Survey. 
}
   {We show that {\it Euclid} will observe hundreds of SLSNe-I for free. These luminous transients will be in the {\it Euclid} data-stream and we should prepare now to identify them as they offer a new probe of the high-redshift Universe for both astrophysics and cosmology.}

\titlerunning{{\it Euclid}: SLSNe in the EDS}
\authorrunning{Inserra et al.}
 \maketitle

\section{Introduction}
Over the last decade, new dedicated transient surveys of the Universe have discovered a multitude of new phenomena. One of the most surprising examples of such new transients is the discovery of 
`superluminous supernovae' \citep[SLSNe,][]{qu11,gy12} which appear to be long-lived explosions (hundreds of days) with peak magnitudes far in excess of normal supernovae \citep[5--100 times the luminosity of Type Ia and core-collapse supernovae,][]{gy12,in13}. 

Over the last five years, it has been established that SLSNe come in different types \citep{gy12,ni15,in16} and can be seen to high redshift \citep[$z\sim 1-4$,][]{co12,ho13}. The power source for these events remains unclear but the most popular explanation is the rapid spin-down of a `magnetar' (a highly magnetic neutron star) which can explain both the peak luminosities and the extended light-curve of SLSNe  \citep{kb10,wo10}. Alternatives include possible interactions between the supernova ejecta and the surrounding circumstellar material previously ejected from the massive central star \citep{chat13}.

With forthcoming surveys like the Zwicky Transient Factory \citep[ZTF;][]{ztf} and the Large Synoptic Survey Telescope \citep[LSST;][]{lsst2,lsst1}, the interest in SLSNe as possible high-redshift cosmological probes has grown due to their high luminosity and possibly increased space density at high redshift \citep{ho13}. Recent studies \citep{is14,pa15,ch16} suggest Type Ic SLSNe \citep[namely hydrogen-poor events with similar spectral features as normal Type Ic supernovae,][]{pa10} could be standardized in their peak luminosities using empirical corrections similar in spirit to those used in the standardization of Type Ia supernova \citep{ru74,ps77,ph93,ham96,rie96,pe97,rie98,go01,guy05,guy07,man09,man11}. \citet{is14} showed that a correction based on the colour of the SLSN-Ic (over 20 to 30 days past peak in the rest-frame) could reduce the scatter in the peak magnitudes to 0.26 \citep[Table 3 in][]{is14} thus raising the possibility that such SLSNe could be used as standard candles. 

This concept was explored in \citet{sco16} where we investigated the potential of SLSNe-I\footnote{Throughout this paper, we will use `SLSNe-I' to refer to Type Ic SLSNe as discussed by \citet{in13}. We do not refer further to Type II SLSNe which appear to have a significantly lower rate than SLSNe-I} for constraining cosmological parameters. \citet{sco16} found that even the addition of $\simeq100$ SLSNe-I to present supernova (SN) samples could significantly improve the cosmological constraints by extending the Hubble diagram into the deceleration epoch of the Universe (i.e. $z>1$). Also, this work predicted that LSST could find $\sim 10^4$ SLSNe-I (over 10 years) which would constrain $\Omega_{\rm m}$ (the density parameter of matter of the Universe) and $w$ (a constant equation-of-state of dark energy) to 2\% and 4\%, respectively. Such a sample of LSST SLSN-I would also provide interesting constraints on $\Omega_{\rm m}$ and $w(z)$ (a varying equation-of-state) that were comparable to that predicted for ESA's {\it Euclid} mission \citep{la11}.

{\it Euclid} is a 1.2m optical and near-infrared (NIR) satellite \citep{la11} designed to probe the dark Universe using measurements of weak gravitational lensing and galaxy clustering. {\it Euclid} is scheduled for launch in late 2020 and will spend the next six years performing two major surveys, namely a `wide' survey of 15,000 deg$^2$ and a `deep' survey of 40 deg$^2$ at both visual (photometry) and NIR (photometry and grism spectroscopy) wavelengths. 

There are proposals to perform a high-redshift Type Ia supernova (SNe Ia) survey with {\it Euclid} \citep[see DESIRE by][]{as14} and WFIRST \citep{wfirst}, which will complement ground-based searches for local and intermediate redshift SNe Ia. DESIRE would be a dedicated 6-month NIR rolling search with {\it Euclid} and is predicted to measure distances to 1700 high-redshift SNe Ia (to $z\simeq1.5$) thus constraining $w$ to an accuracy of 2\%.

In this paper, we study an additional supernova search with {\it Euclid} that is different from DESIRE in two ways. First, we only consider using the already planned {\it Euclid} surveys, specifically the {\it Euclid} Deep Survey (EDS) as it has a planned observing cadence that could be well-suited to the long SLSN light-curves. Secondly, we study the possibility of using SLSNe-I as an additional cosmological probe, which can be seen to higher redshift because of their high luminosities, especially at rest-frame UV wavelengths (although they are not as well-understood as SNe Ia). Therefore, these observations are essentially for free and will be complementary to DESIRE and other {\it Euclid} dark energy constraints. 

In Sect.~\ref{sec:rates}, we outline the rate of SLSN-I as a function of redshift and what is possible with the EDS, while in Sect.~\ref{ss:sp} we give an overview of spectroscopic follow-up of {\it Euclid} SLSNe. In Sect.~\ref{sec:astro} we discuss astrophysical uses of the {\it Euclid} SLSNe, while in Sect.~\ref{sec:cosmo}, we construct a mock Hubble diagram using these {\it Euclid} SLSNe and study the possible cosmological constraints. We discuss {\it Euclid} SLSNe in Sect.~\ref{sec:dis} and conclude in Sect.~\ref{sec:con}. Throughout this paper, we assume a fiducial flat $\Lambda$CDM cosmology with $H_0=68\, {\rm km\, s^{-1}} {\rm Mpc^{-1}}$ and $\Omega_{\rm m}=0.3$, which is consistent with recent cosmological measurements \citep[e.g.][]{au15}.  

\section{Modelling the rate of SLSN-I}\label{sec:rates}

\subsection{The observed SLSN-I rate}
Despite their intrinsic luminosity, there are only approximately 30 well-studied SLSNe-I presently available in the literature with both spectroscopy and multi-band photometric light-curves \citep[e.g. see SLSN-I collections presented in][]{is14,ni15}. However, with forthcoming wide-field imaging surveys (e.g. ZTF, LSST, {\it Euclid}), targeting the distant Universe ($z>1$), we expect the number of such well-studied SLSNe-I to increase significantly over the next decade. 

We focus here on predictions for {\it Euclid}. To make such predictions, we need an estimate of the rate of SLSN-I with redshift. Unfortunately there is still uncertainty in the rate of these rare objects especially at high redshift. For example, \citet{qu13} estimates a SLSN-I rate of 32$^{+77}_{-26}$ yr$^{-1}$ Gpc$^{-3}$ with a weighted mean redshift of $\overline{z} = 0.17$. This corresponds to a fraction ($\sim10^{-4}$) of the volumetric rate of core-collapse SNe (CC-SNe) at the same redshift \citep[consistent with the previous estimate from][]{qu11}. The recent rate measurement of \citet{pr16} using the first four years of the Canada-France Hawaii Telescope Supernova Legacy Survey (SNLS) finds 91$^{+76}_{-36}$ yr$^{-1}$ Gpc$^{-3}$, at a weighted mean redshift of $\overline{z}=1.13$, thus consistent with \citet{qu13}. Between the two measurements there is an increase in the volumetric rate as a function of redshift that is a consequence of the observed star formation history.

However, \citet{mc15} estimate that the SLSN-I rate could be up to ten times lower, based on the Pan-STARRS Medium Deep Survey over the redshift range $0.3 < z < 1.4$, while \citet{co12} obtain an optimistic rate of $\sim200$ yr$^{-1}$ Gpc$^{-3}$ based on only two SLSNe-I at a weighted redshift of $\overline{z}=3.0$. The large uncertainties on all these rate measurements allow them to be consistent with each other, demonstrating that further observations are needed to resolve any apparent discrepancies. 

In addition, if we note that only one SN of the $\simeq50$ Gamma Ray Burst (GRB) SNe appears to be close to superluminous magnitudes \citep{gre15,ka16}, then the rate of SLSN-I is likely to be smaller than the GRB-SN rate by approximately two orders of magnitude. Assuming a ratio of $\simeq4$\% between GRB-SN and SN-Ibc \citep{gd07}, and a rate of $\simeq 2.5 \times 10^{4}$ yr$^{-1}$ Gpc$^{-3}$ for SN-Ibc \citep[from Asiago and Lick surveys,][respectively]{cap99,li11}, the expected rate of SLSN-I would be approximately 10 to 100 objects per yr$^{-1}$ Gpc$^{-3}$, which provides an independent estimate consistent with \citet{qu13} and \citet{pr16}.

\begin{figure*}
\includegraphics[width=18cm]{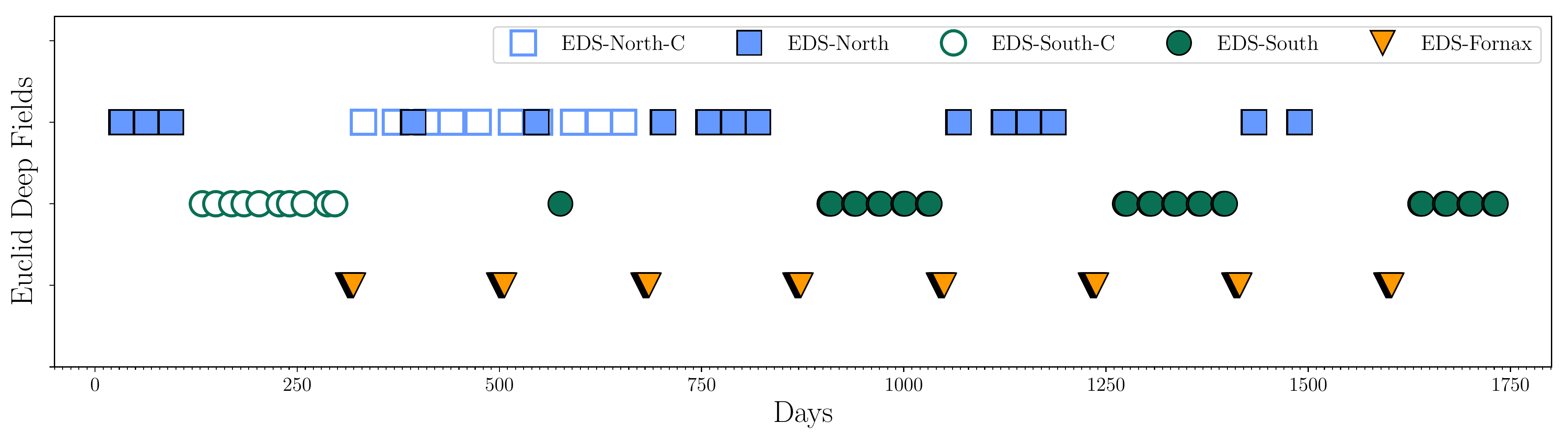}
\caption{Summary of the EDS cadence over the five-year (1825 days) survey. Open symbols refer to the calibration epochs, which are ten per field excluding the Fornax field. Calibration epochs will have the same nominal depth of whole EDS. See Table~\ref{table:cadence} for further details.  }
\label{fig:cad}
\end{figure*}

\begin{table*}
\begin{center} 
\small
\caption{Sampling and coverage information of the three fields of the {\it Euclid} Deep Survey (EDS).\label{table:cadence}}
\begin{tabular}{l|ccccl} 
\hline
Field Name & Area (deg$^2$) & Depth & Visits & Strategy &Additional information \\ \hline
EDS-north & 10 & nominal & 40 &  30 visits to core 10~deg$^2$ (field visited with 2 & north ecliptic pole \\
&  & && consecutive passes) + 10 calibration visits (over 20~deg$^2$) &  \\
EDS-south & 20 & nominal & 40 & 30 visits (clustered every 6 months with 2 consecutive & south ecliptic pole \\
 & &  &  &  passes) + 10 calibration visits. All with 20~deg$^2$ & \\
EDS-Fornax & 10 & nominal & 56 & 7 visits in 7 days every 6 months. All with 10~deg$^2$ & limited visibility in time\\ \hline
\end{tabular}
\end{center}
\end{table*}

\subsection{{\it Euclid} Deep Survey}

To calculate the number of likely {\it Euclid} SLSNe-I, we need to know the volume sampled by the EDS as a function of epoch. The current EDS will likely comprise of three separate areas (see Scaramella et al., in preparation, for further information); one near the north ecliptic pole (EDS-N), one near the south ecliptic pole (EDS-S) and a third overlapping the Chandra Deep Fields South (EDS-Fornax). 

EDS-N is presently scheduled for 40 visits over a five year period. The sampling will not be homogeneous with time differences between consecutive visits ranging from 16 to 55 days (excluding the two 240-day gaps at the beginning and ending of the nominal survey). Ten of these 40 visits will be devoted to calibration purposes (covering an area of 20 deg$^2$), while the remaining 30 visit of EDS-N will scan a central 10 deg$^2$. We only consider this central region in this paper.

EDS-S will also have 40 visits over a five year period, but will cover an area of 20 deg$^2$. These observations will be clustered in six-month blocks with an average cadence between visits of 28 days. Every 28 days two visits will be grouped in a three-day window. Discover astronomical transients in a field rich of foreground stars will not be a problem if algorithms using supervized machine learning techniques are employed as done by current transient surveys \citep[e.g.][]{bloom12,goldstein15,wright15}.

EDS-Fornax (covering an area of 10 deg$^2$) will be observed 56 times, to compensate for the expected higher background, with a limited visibility. It will be observed every day for a week with gaps of six months between the week of visibility. 

We present a summary of these three EDS fields in Fig.~\ref{fig:cad} and Table~\ref{table:cadence}. For this work, we have ignored the EDS-Fornax because of its low-visibility and therefore, the final areal coverage assumed is 30 deg$^2$ over the first two fields (north central area plus the whole southern area).

\begin{figure}
\includegraphics[width=9cm]{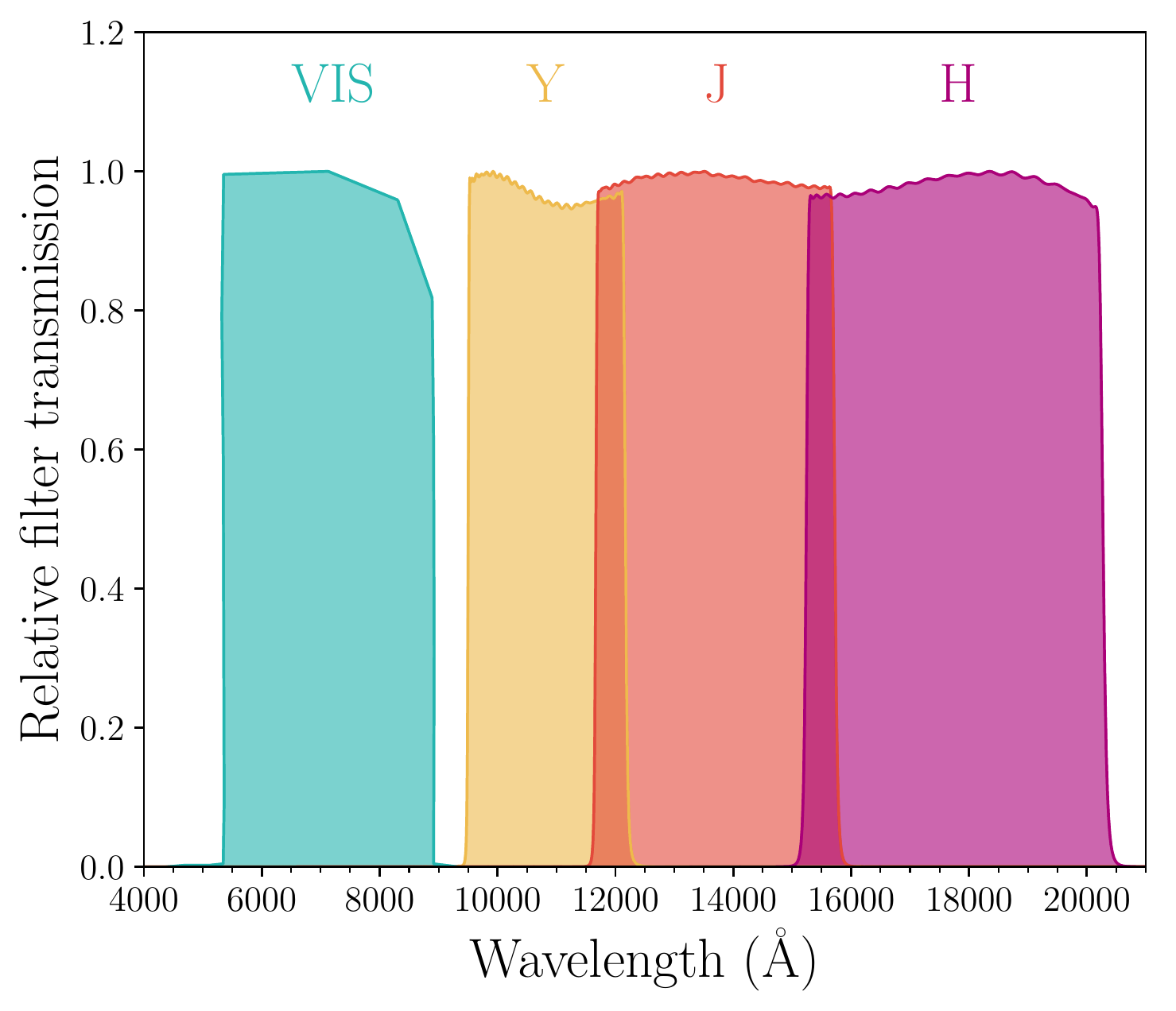}
\caption{Normalized filter transmission of $VIS$ and $Y$, $J$, $H$ (NISP).}
\label{fig:filter}
\end{figure}

\begin{table}
\begin{center} 
\small
\caption{{\it Euclid} filters specifications.\label{table:filters}}
\begin{tabular}{c|cc} 
\hline
Filter Name & Central wavelength (\AA) & Width (\AA) \\ 
\hline
$VIS$ & 7150 & 3550 \\
$Y$ & 10850 & 2750\\
$J$ & 13750 & 4300 \\
$H$ & 17725 & 5250\\
\hline
\end{tabular}
\end{center}
\end{table}

We assumed a 5$\sigma$ limiting magnitude of 25.5 for each of the individual EDS visual visits ($VIS$ passband is equivalent to $r$+$i$+$z$ passbands, see Fig.~\ref{fig:filter} and Table~\ref{table:filters}), while we assumed $Y=J=H=24.05$ magnitudes for each Near Infrared Spectrometer and Photometer (NISP) visit of the EDS. These point source values are slightly different from those reported in \citet{as14}, but are in agreement with the {\it Euclid} science requirements \citep[][]{la11} and the latest estimates of the {\it Euclid} performance. We assumed the latest filter transmission curves and quantum efficiencies for $VIS$ \citep[][]{cro14} and NISP. The filter transmission functions are shown in Fig.~\ref{fig:filter}, and have been implemented in the {\tt S3} software package \citep[see][for further details on the programmes]{in16} for $k$-corrections. We note that with such a steady cadence, and consistent limit magnitudes per visit, there will always be at least ten epochs for each detected supernova (at any redshift) that can be used as a reference image (e.g. without SN light) for difference imaging.

\begin{figure*}
\includegraphics[width=18cm]{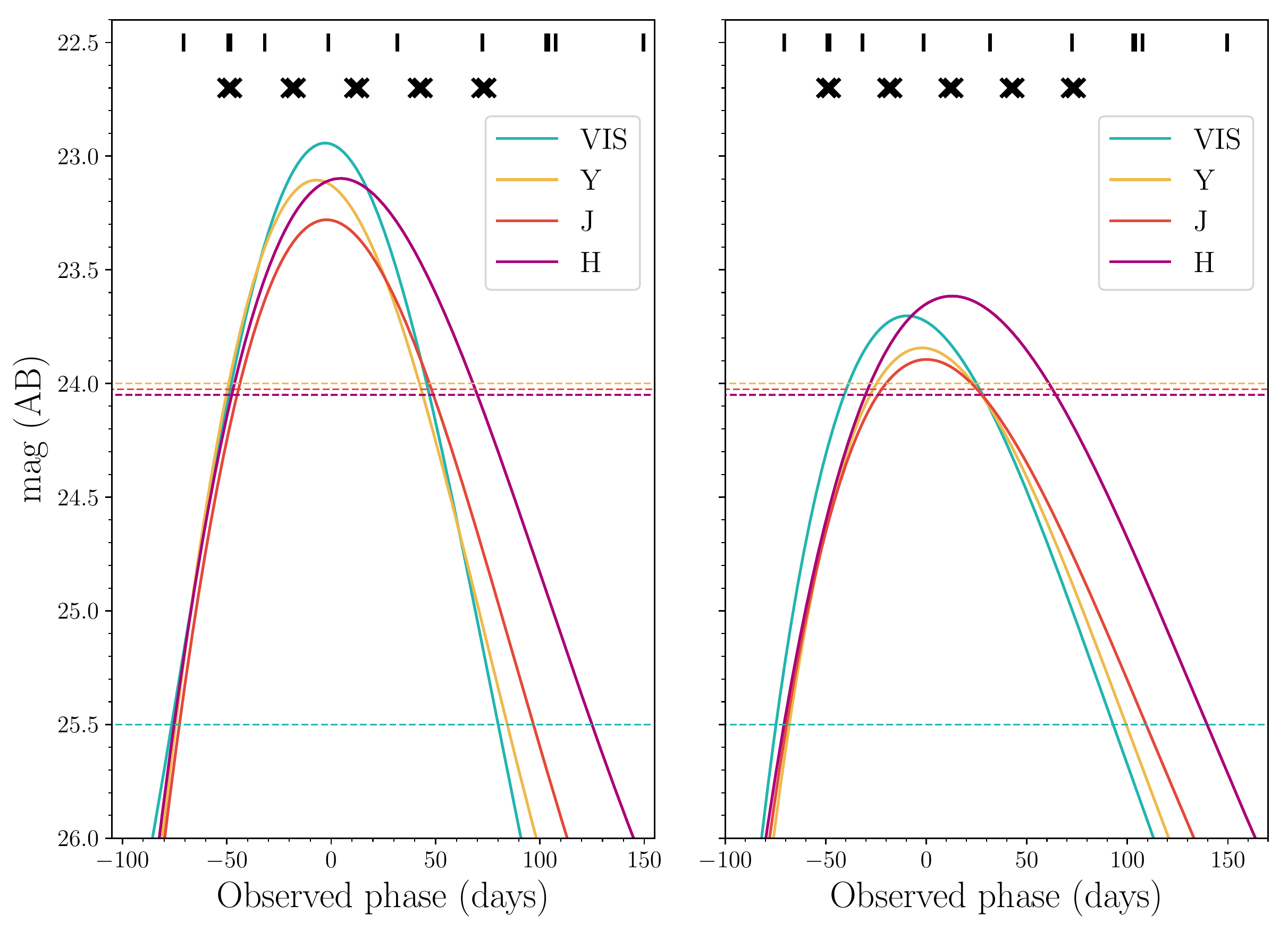}
\caption{Left: Simulated observer-frame light-curve for a $z=2.0$ SLSN-I in the four {\it Euclid} passbands. The horizontal (dashed) lines represent the assumed 5$\sigma$ point source limiting magnitudes for each filter as discussed in the text. $J$ and $H$ limiting magnitudes are shifted of $0.05$ and $0.10$ magnitudes to facilitate the reading. The cross symbols at the top of the panel represent a typical observing cadence for the southern EDS away from the six months gap (including two consecutive observations per passage that are not considered in the rate simulations), which would detect this SLSN four times in three bands. Similarly, the short lines at the top of the panel represent a typical observing cadence for the northern EDS, with four detections in three bands. Right: The same as the left panel but at $z=3.5$. In this case both the southern and northern EDS would detect this SLSN three separate times (again excluding double observations within three days of each other). Observed phase is with respect to the observer frame $J$-band peak.}
\label{fig:lc}
\end{figure*}

\subsection{Luminosity function}\label{ss:lfun}

We adopted a luminosity function with an average light-curve peak of $-21.60 \pm0.26$ $r$-band magnitude, rising for $25\pm5$ days and declining $1.5\pm0.3$ mag in 30 days \citep{is14,ni15}. This has been built fitting literature data \citep[e.g.][]{in13,ni15} with low-order polynomials \citep[as done in][]{ni16} and the magnetar model \citep{in13}, allowing a reduced $\chi^2\lesssim5$. Such a luminosity function is in agreement, within the uncertainties, with the recent findings of \citet{decia17} and \citet{lunnan17}.
We utilized an empirical template for the spectral energy distribution (SED) of SLSN-I based on 110 rest-frame spectra taken for 20 SLSNe-I spanning a redshift range of $0.1<z<1.2$ (from 1800~\AA\/ to 8700~\AA) and covering approximately $-20$ to 250 days (with respect to peak luminosity) in their light-curve evolution \citep{gy09,pa10,qu11,in13,ni13,ni14,vr14,ni16}. 

This template was implemented in the {\tt snake} software package \citep{in16} to calculate the necessary $k-$corrections between the assumed {\it Euclid} visual and NIR filters and the standard optical rest-frame passbands, namely the SDSS $r$-band and the two narrow passbands used in \citet{is14} to standardize SLSNe-I (namely, their 4000 and 5200~\AA\ synthetic filters). We use the SDSS $r$ filter as our main reference rest-frame filter for our calculation. Uncertainties on the $k-$corrections have been evaluated as RMS of the uncertainties on redshift, spectral template and different standard passbands following the methodology of \citet{in16}. Usually such uncertainties are smaller than $0.05$ magnitudes \citep{kim96,br07,hs07,in16}. Assuming that the terms leading to the definition of our observed magnitude ($m$) are uncorrelated, and the uncertainties deriving from the cosmology adopted are negligible, these uncertainties are given by

\begin{equation}
\begin{split}
& \sigma (m)^{2}  =  \sigma (M)^2 - 4.7\left(\frac{\sigma (D_{\rm L})}{D_{\rm L}}\right)^2  -\sigma (A)^2 -1.2 \times \\
& \left[ \left( \frac{\sigma (z)}{z}\right)^2 + \left( \frac{\sigma (L_{\lambda_{\rm o}})}{L_{\lambda_{\rm o}}}\right)^2 + \left( \frac{\sigma (L_{\lambda_{\rm r}})}{L_{\lambda_{\rm r}}}\right)^2 + \left( \frac{\sigma (ZP_{\lambda_{\rm o}})}{ZP_{\lambda_{\rm o}}}\right)^2 + \left( \frac{\sigma (ZP_{\lambda_{\rm r}})}{ZP_{\lambda_{\rm r}}}\right)^2 \right] \: 
\end{split}
,\end{equation}

\noindent where $M$ refers to the absolute magnitude, $D_{\rm L}$ is the luminosity distance, $A$ is the extinction coefficient, $L_{\lambda}$ is the luminosity function in that filter, $z$ the redshift, $ZP$ are the filter zero-points, and $o$ and $r$ refer to the observer and rest-frames, respectively \citep[see][]{is14,in16}.  All these uncertainties are included in the uncertainties estimate, with the exception of the uncertainties on the host galaxy extinction, which are usually negligible for SLSNe-I \citep{is14,ni15,le15}.

\begin{figure*}
\includegraphics[width=18cm]{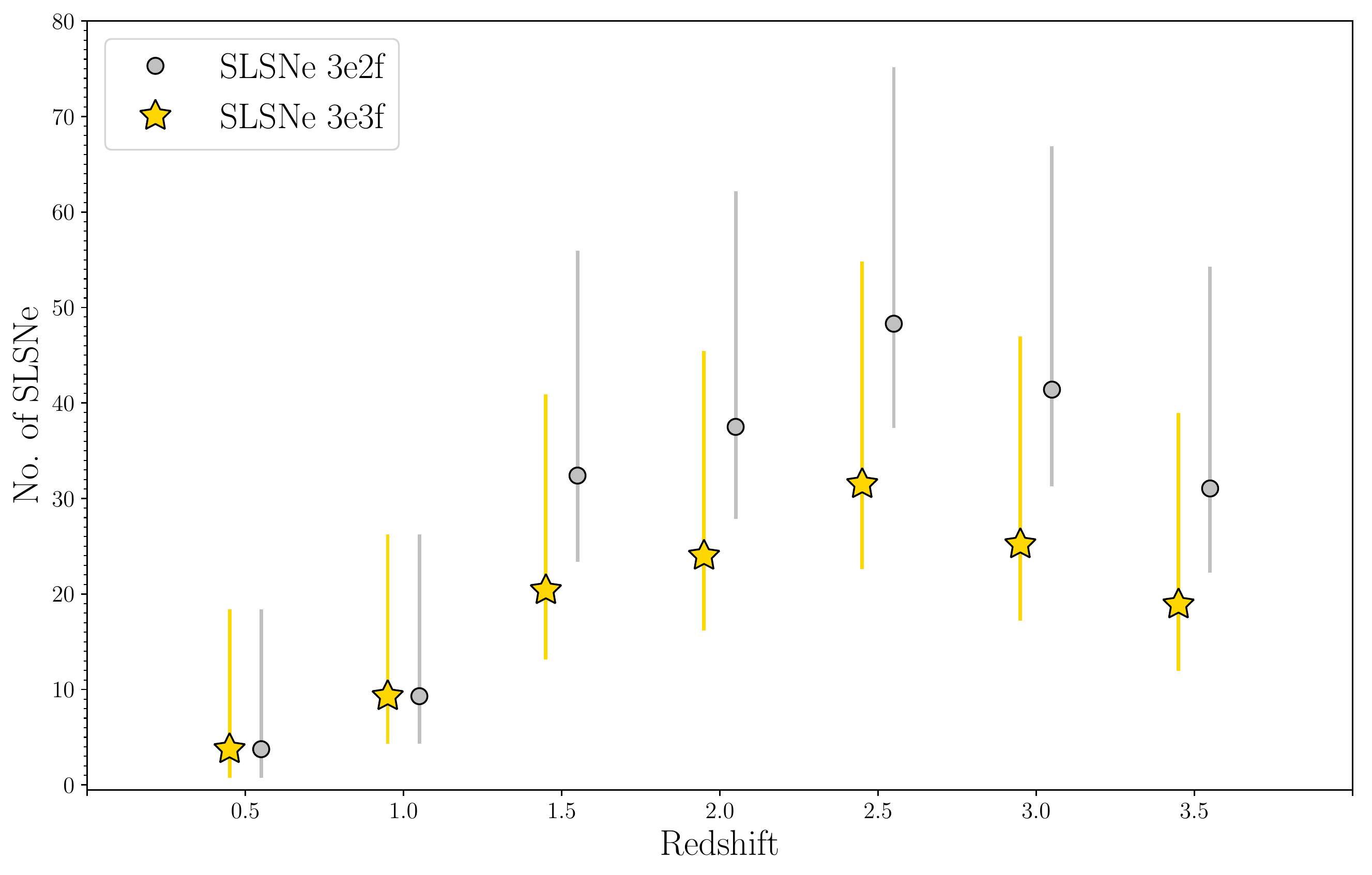}
\caption{Number of SLSNe-I detected, per redshift bin ($\Delta z=0.5$), during the five years of the EDS (combining both the northern and southern EDS observations). Gold stars denote the `gold sample' (three filter detections for each of three epochs, or 3e3f in legend), while the silver circles are the `silver sample' (two filter detections for each of three epochs, or 3e2f). The error bars are Poisson uncertainties based on the number of SLSNe-I in each bin \citep{ge86}, while the rates assumed are for our optimistic model (see text). Both gold and silver points are offset of $\Delta z=0.05$ to facilitate the reading.}
\label{fig:rates}
\end{figure*}

\subsection{{\it Euclid} SLSN-I rate}

In order to estimate the volumetric rate of SLSN-I in the EDS, we used a model for the evolution of the star-formation rate (SFR) density with redshift \citep[see][]{hb06}, based on the Salpeter initial mass function (IMF) published by \citet{co01}, and using the methodology of \citet{bot08}. Adopting an IMF and SFR at any redshift then allowed a calculation of the volumetric rate of core-collapse supernova, assuming that all stars above $8~M_{\odot}$ produce a SN (the upper mass limit is not important as long as its $\gtrsim50~M_{\odot}$). We then assumed that the ratio between the SLSN-I and CC-SN rate is $10^{-4}$ \citep{qu13,pr16}, which provides a rate per co-moving volume element. We will refer to this rate as the `optimistic' model. 

To estimate the systematic uncertainty on this rate, we also re-calculate it adopting a slightly different evolution for the SFR density \citep{bo11} as well as the lower ratio of $10^{-5}$ between the SLSN-I to CC-SN rates \citep{mc15}. This will be our `pessimistic' model. This approach gives more freedom and allowed us to have better uncertainties in case SLSNe do not follow the SFR of the bulk of the Universe. In fact low-metallicity, faint, galaxies appear to have a much flatter SFR with redshift than the nominal SFR of the Universe \citep[see][]{he04}.
We assume Poisson statistical uncertainties on both estimates. We note that the optimistic set-up is the one consistent with other SLSN-I rate estimates up to redshift $z\sim1$ \citep[see Fig.~9 of][for a comparison]{pr16} and those at higher redshift \citep{ta12,ta13}. SLSNe-I host galaxies properties, such as metallicity, star formation rate and stellar mass, do not show any obvious redshift dependence \citep[][]{lu14,le15,ch16} but only a general metallicity threshold (12 + log(O/H)$_{\rm N2}<8.5$), which however strongly depends on the diagnostic used \citep[see][for an in-depth discussion]{ch17}. Hence, we do not expect any significant quantitative change in our assumptions (SFR density and SLSN-I to CC-SN rate) with redshift.

These two models are then used to calculate the number of SLSNe-I, in bins of $\Delta z=0.5$ width, centred on multiples of $z=0.5$ up to $z=3.5$ which is consistent with the highest SLSN-I redshift observed to date, and likely achievable with {\it Euclid} \citep[see][]{is14}.

We then performed $10^5$ Monte Carlo simulations\footnote{This level of simulations has been used in previous literature studies \citep[e.g.][]{pr16} and found to be adequate.} of {\it Euclid} SLSN-I light-curves for each bin of $\Delta z=0.5$ and placing them at random explosion epochs relative to the EDS observing strategy that is, the survey time, depth, cadence and volume as discussed above.  We show in Fig.~\ref{fig:lc} two examples of such simulated light-curves at $z=2.0$ and $z=3.5$. During these simulations, we also assumed an average foreground extinction of $E(B-V)=0.02$ \citep[see][]{is14}.

\begin{table*}
\caption{\label{table:rates} Number of SLSNe-I per year for both samples (silver and gold) and with both rate models (see text). These data are shown in Fig.~\ref{fig:rates}. The redshift bin width is $\Delta z=0.5.$}
\centering
\begin{tabular}{llccccccc}
\hline\hline
\multicolumn{9}{c}{Optimistic}\\
\hline
Years & criteria & 0.5 & 1.0 & 1.5 & 2.0 & 2.5 & 3.0 & 3.5  \\
\hline
1-5 & silver & 1  & 2  &  6  & 8  & 10 & 8 &  6 \\
1-5 & gold & 1 &  2  &   4  & 5  &  6   & 5  & 4  \\
\hline\hline
\multicolumn{9}{c}{Pessimistic}\\
\hline
Years & criteria & 0.5 & 1.0 & 1.5 & 2.0 & 2.5 & 3.0 & 3.5 \\
\hline
1-5 & silver & 1  & 1  &  4  & 5  & 6 & 5 &  4 \\
1-5 & gold & 1 &  1  &   2  & 3  &  4   & 3  & 2  \\
\hline
\end{tabular}
\end{table*}

Using these simulated light-curves, we then determined which SLSNe-I would be useful for any meaningful astrophysical and cosmological analysis. We therefore defined two sub-sets of SLSN-I using the following selection criteria. First, we defined a `silver sample' that requires each SLSN to be detected (5$\sigma$ point source) for at least three epochs (3e) in their light-curves in at least two {\it Euclid} filters (2f) per epoch (or 3e2f). Second, we defined a `gold sample' which requires a detection (5$\sigma$ point source) in at least three {\it Euclid} filters, each for at least three epoch (3e3f). In all cases, we required at least one of these detections to be before peak brightness. In addition, we only considered epochs that are separated by at least three days to ensure reasonable coverage of the whole light-curve (Fig.~\ref{fig:lc}). We simply ignored all but one epoch of those more closely separated by less than three days. These extra (close) epochs would not provide any additional information in terms of the light curve sampling, and colour evolution at $z\gtrsim2$, but in reality they could be helpful for SN detection (e.g. removing bogus artifacts, asteroids, and cataclysmic variables) and increased signal to noise ratio (S/N). 

Moreover, these close epochs ($\Delta t<3$ days) in both EDS-S and EDS-N would be excellent for discovering fast transients such as rapidly evolving SNe \citep{dr14} or observe red kilonovae \citep[e.g.][]{ka15,met15a}. The latter are fast transients visible for approximately two weeks as a result of two neutron stars merging and likely producing gravitational waves in the sensitivity region of the LIGO interferometers \citep[strain noise amplitudes below 10$^{-23}$ Hz$^{-1/2}$ in the frequency regime 10$^2$--10$^3$ Hz;][]{ber15,ligo}.

These criteria should be sufficient to efficiently separate SLSNe-I from other transients\footnote{Dark Energy Survey, private communication. Angus et al. in preparation}, such as active galactic nuclei and high-$z$ lensed SNe Ia because of their characteristic photometric colour evolution \citep[see the colour evolution shown by][]{in13,is14,ni15}. Also, recent work has shown that the evolution of the luminosity and colour of SLSNe trace a distinctive path through parameter space \citep[see][]{in17b}, while targeting apparently 'hostless' SLSN candidates can also improve the success rate of any spectroscopic follow-up \citep[e.g.][]{mc15}. Finally, SLSNe have been shown to possess similar spectrophotometric evolution up to $z\sim2$ \citep{pan17, smi17}, which supports our assumption on the luminosity function and our analysis in Sect.~\ref{sec:cosmo}. Furthermore, requiring a detection in at least two passbands will provide at least one colour measurement which is essential for using the relationship between peak magnitude and colour evolution as discussed in \citet{is14} for standardization. Having three passbands would provide a better estimate of the bolometric light-curve which could further be used to standardize SLSNe-I, for example correlating the spin period of the best-fit magnetar model to the host galaxies metallicity  \citep[see][]{ch16}. 

In Fig.~\ref{fig:rates},  we show the results of our simulation. When determining the number of SLSNe-I expected from the EDS, we assume that only the northern and southern areas of the EDS (total of 30 deg$^2$) are observed as shown in Fig.~\ref{fig:cad}. In the case of our optimistic model, this provides a yearly volumetric rate of $41^{+11}_{-6}$ yr$^{-1}$ Gpc$^{-3}$ for the silver sample (3e2f) and $27^{+9}_{-4}$ yr$^{-1}$ Gpc$^{-3}$ for the gold sample (3e3f). 
Uncertainties are Poisson and have been estimated using \citet{ge86} for small numbers of events in astrophysics. In addition, uncertainties on the yearly rates at $0.5<z<3.5$ have been estimated with the same formalism applied to the sum of each bin since the sum of each independent Poisson random variable is Poisson.

We present in Table~\ref{table:rates} the expected number of SLSNe-I as a function of redshift for both rates models, while in Fig.~\ref{fig:rates} we only plot the results of the optimistic model. In total, we predict {\it Euclid} will detect $\simeq 140$ high-quality (gold sample) SLSNe-I up to $z \sim 3.5$ over the five years of the EDS. On the other hand, the silver sample could deliver an extra 70 SLSNe-I, with respect to the gold, over the same five years. Extending the EDS beyond the nominal five-year duration could add approximately 40 SLSNe-I per year to the sample. If in our optimistic model we increase the detection level from 5$\sigma$ to 10$\sigma$, then the predicted rates would be similar to those for the 5$\sigma$ pessimistic model. Furthermore, if the limiting magnitude of the $VIS$ filter was less efficient than assumed here, and closer to the initial 24.5 magnitude limit originally expected, we would see a drop of $\sim7$\% in our rate.

\section{SLSN Spectroscopy}\label{ss:sp}

Throughout this analysis, we have assumed we know accurately the redshift and identification of the detected SLSN-I events. Such information could be achieved through fitting models of SLSN to the {\it Euclid} photometric data as previously done with SNe Ia \citep[e.g.][]{jha07,guy10,su11} or fitting the magnetar model to the multi-colour light-curve \citep[e.g.][]{pr16}. However, these approaches are model-dependent and there are degeneracies in such fitting techniques leading to systematic biases.

The logical next step is to secure a spectrum, especially after peak \citep[see][]{in17b} for as many of these SLSNe-I as possible to determine both their redshift and classification. This has traditionally been the approach for SNe Ia, but recently the number of detected SN Ia candidates is far beyond our capability to perform real-time spectroscopic follow-up of all these events \citep[see][]{2013ApJ...763...88C}. For example, the Dark Energy Survey (DES) is now focussed on gaining spectroscopic follow-up for all SN host galaxies which is easier given large multi-object spectrograph \citep[see][]{yuan15}. However, such an approach may not work for all SLSNe-I as many of these events happen in low-mass, compact dwarf galaxies \citep{lu14}.

Fortunately, the rate of new {\it Euclid} SLSN detections should be approximately once a week, and each supernova will last for several months in the observer-frame (see Fig.~\ref{fig:lc}). Therefore, it should be feasible to obtain real-time spectroscopic follow-up of these {\it Euclid} SLSNe, unlike the LSST SLSN sample where we may find $\sim25$ new SLSNe-I per week \citep{sco16} over the ten years of operations. Also, the intrinsic rate of SLSN-I events should be far in excess of the expected SLSN-II rate, meaning the expected contamination from such events (e.g. for our cosmological analysis) will be minimal, although we note the discovery of more high-redshift SLSNe-II would be of great interest for astrophysical studies. 

\begin{figure}
\includegraphics[width=\columnwidth]{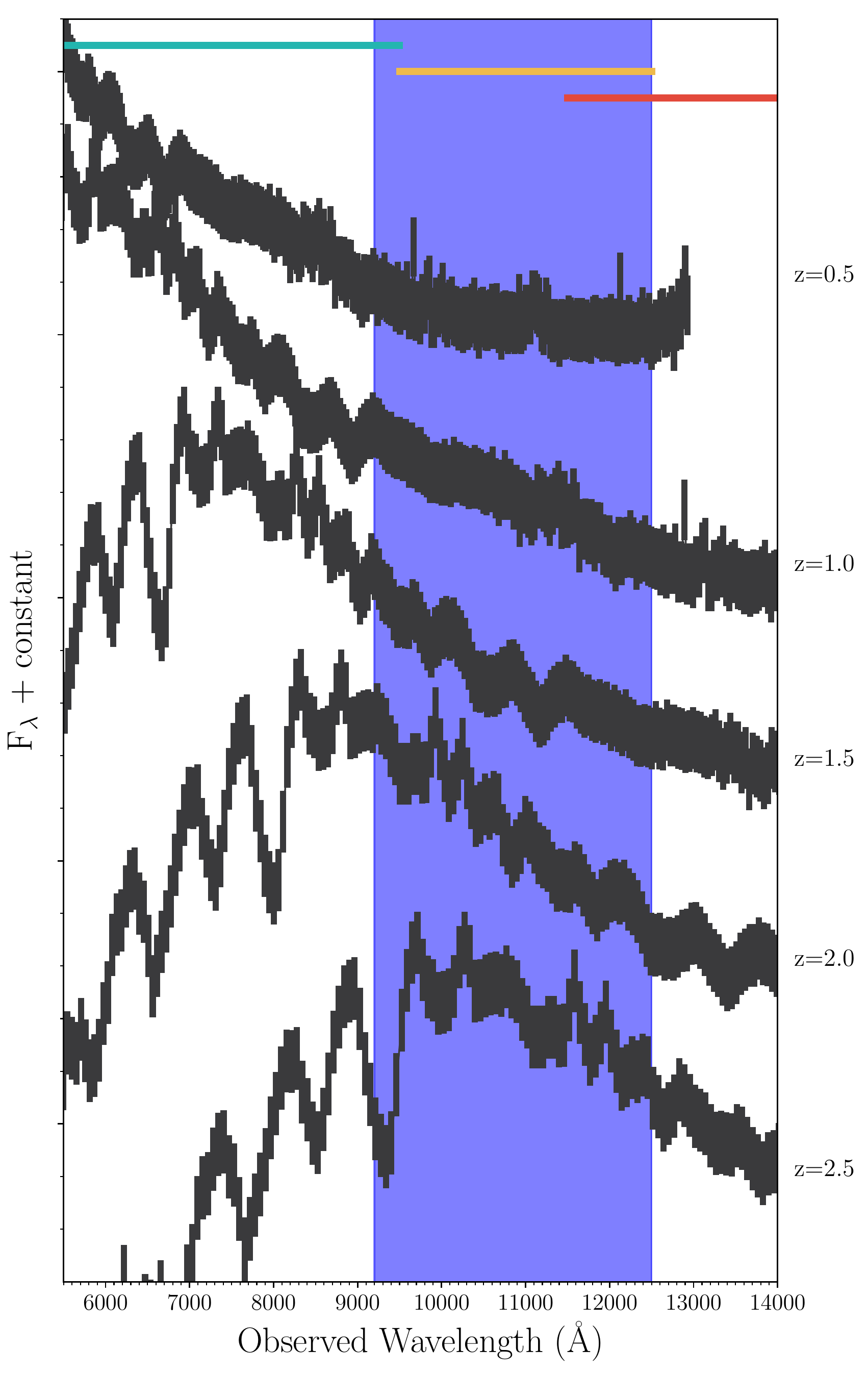}
\caption{Example of an observed spectrum with ${\rm S/N} \sim20$, $R\simeq350$ of a SLSN-I at various redshifts ($0.5<z<2.5$). The spectrum is that of iPTF13ajg at peak epoch \citep{vr14} with a flux of $\sim 10^{-16}$ erg s$^{-1}$ cm$^{-2}$ \AA$^{-1}$ at $Y$-band wavelength and at redshift $z=0.5$, which should be feasible for the {\it Euclid} `blue' grism (see text).  The cyan, yellow and red solid lines represent the wavelength regions covered by {\it Euclid} $VIS$, $Y$ and $J$ filters, respectively (we note that $VIS$ and $Y$ are superimposed for 600~\AA). The blue region shows the {\it Euclid} `blue' grism covering $0.92$ to  $1.25$ microns. The spectra at $z>0.5$ show the potential of future facilities (e.g. JWST, E-ELT) and their use in identifying SN features, since they will go deeper and with a better resolution than the {\it Euclid} spectrograph.}
\label{fig:sp}
\end{figure}

To assess the feasibility of obtaining spectra of these {\it Euclid} SLSNe-I, we use the exposure time simulator of the near-infrared integral-field spectrograph HARMONI \citep{hsim} planned for the European Extremely Large Telescope (E-ELT). We  estimate that an effective exposure time of 900 seconds would give 
a S/N of $\sim$20 for an average SLSN-I at $z=2$. Such S/N is sufficient to identify a transient as shown in Fig.~\ref{fig:sp} and by the Public ESO Spectroscopic Survey of Transient Objects \citep[PESSTO,][]{pessto}. Alternatively, it would only require 300 seconds to achieve a similar S/N for the same SLSN-I using the low-resolution grating on the Near-Infrared Spectrograph (NIRSpec) on the James Webb Space Telescope (JWST). We anticipate similar exposure times for the Wide Field Infrared Survey Telescope (WFIRST), as well as the 30\,m Thirty Meter Telescope (TMT) and the 23m Giant Magellan Telescope (GMT). 

On the other hand, it may be more challenging with existing ground-based eight-meter telescopes. For example, we would require a two-hour integration using the Near-Infrared Integral Field Spectrometer (NIFS) on the Gemini telescope, to achieve ${\rm S/N} \sim 5$, which is a lower limit for identifying transients object with broad feature like SLSNe-I and secure a redshift. 

It may also be possible to obtain some spectral information for these SLSNe-I from the low-resolution ($R\simeq350$) {\it Euclid} NIR slitless spectroscopy that is planned for the EDS in parallel to the imaging data. The information on the NIR slitless spectroscopy here reported is the latest available to the {\it Euclid} consortium and likely to be the final, even though we warn the readers that later changes could always happen. The {\it Euclid} NISP instrument has two NIR grisms, namely a `blue' grism covering $0.92$ to  $1.25$ microns and a `red' grism covering wavelengths of $1.25$ to $1.85$ microns. The EDS slitless spectroscopy strategy may focus primarily on observations with the blue grism with approximately three-quarters of the deep field visits (each of 4 dithers) dedicated to this grism. We expect each visit to reach a limiting flux of $2\times 10^{-16}$ erg s$^{-1}$ cm$^{-2}$ \AA$^{-1}$ (3.5$\sigma$) to match the depth of the {\it Euclid} wide survey (for calibration purposes). We estimate that such data may provide spectroscopic information on a live SLSN (as detected in the imaging) to $z\leq0.5$. Better performance could be obtained by re-binning the spectrum as the SLSN-I features are broad and optimally extracting the spectral data using our prior knowledge of a candidate SLSN at that location. In addition, in case of nearby SLSNe, a spectrum observed after peak should contain more information as both the O~{\sc i} and Ca NIR lines will be present in the observed region of a blue NIR grism configuration.  

We may also obtain the redshift of a SLSN-I host galaxy through the co-addition of multiple NISP grism spectra, at the known location of the SLSN, to achieve a possible flux limit of $3\times 10^{-17}$ erg s$^{-1}$ cm$^{-2}$ \AA$^{-1}$ (over a one arcsecond aperture). Many SLSNe-I are located in star-forming dwarf galaxies with strong oxygen nebular emission lines that means we should detect [O {\sc iii}] to $z=1$ in the blue grism for several of the SLSN-I host galaxies \citep[e.g.][]{le15,pe16}.

\section{Astrophysics from high-redshift SLSNe}\label{sec:astro}
The discovery of hundreds of SLSNe-I in the EDS will improve several areas of astrophysics. For example, {\it Euclid} photometry and spectroscopy of nearby SLSNe-I will improve our knowledge of their SEDs in the NIR (only up to $z=0.5$ objects), where the uncertainties due to extinction are minimized. This would then allow us to compare SLSNe-I with similar photospheric spectra, but different observed colours and continuum slopes, to gain insights into the host galaxy extinction. On the other hand, Euclid will deliver hundreds of SLSNe over a longer redshift baseline than those currently available. This will increase the statistical power providing for a more principled approach in the spectrophotometric analysis. Consequently, it could lead to a better understanding of the mechanism responsible for the luminosity of SLSNe-I, which has been narrowed to an inner engine, spin down of a rapidly rotating magnetar \citep[e.g.][]{kb10,wo10,de12} or a black hole \citep{dk13} and/or interaction of the SN ejecta with a massive ($3-5$ M$_{\odot}$) C/O-rich circumstellar medium \citep[e.g.][]{wo07,chat13}. We also note that a pair instability explosion \citep[e.g.][]{kb15} could still be a viable alternative at high redshift after having been disfavoured for SLSNe-I at $z\lesssim 1.5$ \citep[e.g.][]{ni13,smi16,in16b,in17a}.

Due to their high luminosity, SLSNe-I are also excellent probes of the physical conditions of the gas surrounding the SN, as well as the interstellar medium within the host galaxy interstellar medium. This is possible through the detection of broad UV-absorption lines, which will be redshifted into the NIR wavelength range. For example, elements like Mg, Si, Fe, and Zn can be detected, via narrow absorption lines in the follow-up spectra of SLSN-I, thus allowing us to measure the metal column densities, relative abundances, dust content, ionization state and kinematics of the gas. First attempts to detect such metal lines in the rest-frame UV of such SLSNe-I have been successful \citep[e.g.][]{be12,vr14}. Furthermore, any Fe and Ni within 100 parsecs of the SN should be excited via `UV-pumping', thus providing an estimate of the distance between the SN and any absorbing gas \citep[as achieved for GRBs, e.g.][]{vr13}. This, combined with the velocity information, would provide a novel constraint on the immediate environment, and progenitors, of these SLSNe-I.

In Sect.~\ref{sec:rates}, we estimated the number of SLSNe-I observed by {\it Euclid} by assuming they follow the cosmic star-formation history, since these events are proposed to originate from massive stars \citep[e.g.][]{je16}. Such an assumption will be tested via these {\it Euclid} SLSNe-I, especially with those objects beyond $z>1.5$, which is currently the limit of the reliability of rate estimates \citep{pr16}. SLSNe-I are associated with low-metallicity, high star-forming galaxies \citep[e.g.][]{lu14,le15,ch16} at all observed redshifts \citep[$z<4$;][]{co12}, hence we do not expect that to change at {\it Euclid} SLSNe-I redshifts. This implies that {\it Euclid} SLSNe-I will also trace the cosmic chemical enrichment as previously done with CC-SNe \citep[e.g.][]{2015ApJ...813...93S}.

Furthermore, this large number of {\it Euclid} SLSNe-I discovered over a wide redshift range will improve our understanding of stellar explosions and transient events, for example, {\it Euclid} could discover interesting objects similar to SN~2011kl \citep{gre15,ka16,be16} and ASASSN-15lh \citep[e.g.][]{do16,le16,vd17,mar17} which achieve similar luminosities as SLSNe, but show different spectrophotometric evolution. 

\section{SLSN cosmology}\label{sec:cosmo}
\subsection{Methodology}
Following the work of \citet{sco16}, we can also consider the cosmological usefulness of the {\it Euclid} SLSNe-I. For this analysis we explored what could be achieved if we were to obtain a sample of 300 {\it Euclid} SLSNe-I with the redshift distribution given in Fig.~\ref{fig:rates}. Such an optimistic sample may be possible if we can utilize the additional silver sample ($140+70$) and obtain an extension to the EDS beyond the nominal five-year duration of the {\it Euclid} mission, for example like DESIRE. Moreover, we may find even more SLSNe-I given the present uncertainties in the high-redshift SLSN rate, and our assumed luminosity function (which is quite conservative). 

We note that, despite such SLSNe-I showing a luminosity function with a gaussian-like distribution \citep[e.g.][]{is14,ni15,in17b,lunnan17}, current surveys are starting to populate the lower luminosity end creating a continuum in luminosity between normal and superluminous SNe \citep[see][]{decia17}.

We followed the methodology outlined in \citet{sco16} to construct a mock Hubble diagram (redshift-distance relationship) for such a sample. As in \citet{sco16},  we include the additional magnitude dispersion of weak gravitational lensing, which will be important for high-redshift objects for example beyond $z\simeq2$ \citep{marra2013}. We also assume that all the SLSNe-I have been successfully classified (see Sect.~\ref{ss:sp} for discussion of spectroscopic follow-up of these events) and our sample contains negligible contamination (e.g. outliers on the Hubble diagram). 

In addition to the high-redshift SLSNe-I, we need to include a low redshift sample to help anchor the Hubble diagram. Therefore, we assumed 50 SLSNe-I, homogeneously distributed over the redshift range $0.1<z<0.5$ for this local sample. This choice is consistent with \citet{sco16} and is rather conservative given existing samples of low-redshift SLSN-I in the literature and expectations from planned, and on-going, transient searches like ASASSN \citep{2014ApJ...788...48S} and ZTF, and spectroscopic follow-up programmes like PESSTO \& ePESSTO \citep[][]{pessto}. 
Within each redshift bin, we assigned the redshift value at random for the number of supernovae given in Table~\ref{table:rates}. 

We combined the SLSN mock Hubble diagram with a DES \citep[see][]{Bern12} mock sample for SN Ia\footnote{Details of this mock sample can be found in Sect. 3.2 of \citet{sco16}}. It is composed of 3500 SNe Ia \citep[distributed according to the hybrid-10 strategy in][]{Bern12} and 300 low-redshift SNe Ia (uniformly distributed for $z<0.1$).

We fitted our mock Hubble diagram using the publicly available code {\tt cosmomc} \citep[July 2014 version,][]{cosmomc}, run as a generic Markov chain Monte Carlo sampler. This allows us to include a custom-made likelihood in the software,
\begin{equation}
\label{eq:liks}
\mathcal{L}=\mathcal{L}_{\rm SLSN}\times \mathcal{L}_{\rm SNIa},
\end{equation}
defined as the product of two likelihoods, one for each of the data samples  considered here (`SLSN' and `SNIa' hereafter).

Both of these likelihoods have the same functional form,
\begin{equation}
\mathcal{L}=\frac{1}{(2\pi)^{n/2}\sqrt{\det C}}\exp \left[-\frac{1}{2} \left(\boldsymbol{\Delta \mu}^T C^{-1} \boldsymbol{\Delta \mu} \right) \right],
\end{equation}
where $\boldsymbol{\Delta \mu}$ is the $n$-dimensional vector containing the Hubble residuals (see below) and $n$ is the number of supernovae in that sample. As discussed in \citet{sco16}, we neglect the covariance between supernovae (i.e. all the non-diagonal terms are set to be zero) as we expect these to be small compared to the statistical noise of the limited sample sizes and gravitational lensing (see below). We also do not yet have a good understanding of possible systematic uncertainties (e.g. the photometric calibration) but assume they will be sub-dominant given present expertise in calibrating such photometric surveys. Also, the science requirement on the {\it Euclid} relative photometry is 0.002 magnitudes, which exceeds the calibration uncertainties with present ground-based large-area surveys.  

Hence, each covariance matrix $C$ would reduce to diagonal elements only, giving
\begin{equation}
 C_{ij}=\langle\Delta\mu_i \Delta\mu_j \rangle = \sigma^2 _{ij}\delta_{ij}=\sigma_{err}^2\delta_{ij}+\sigma_{\rm len}^2(z_i)\delta_{ij}.
\end{equation}
Each measurement then has an uncertainty equal to the sum in quadrature of the data and lensing uncertainties \citep[respectively $\sigma_{\rm err}$ and $\sigma_{\rm len}(z_i)$, see Sect. 3.2 of][]{sco16}. We assumed that the magnitudes of the SLSN-I population will be standardized using techniques like those outlined in \citet{is14}, or more advanced future techniques similar to those now used for SNe Ia \citep[e.g. SALT and BayeSN;][]{guy10,man11}. To be conservative, for our SLSN-I mock samples, we assumed a dispersion $\sigma_{\rm err}=0.26$ magnitudes, based on the findings of \citet{is14} and following the previous work of \citet{sco16}. Specifically, we select this value for $\sigma_{\rm err}$ from Table 3 of \citet{is14} based on their $\Delta(400-520)$ extended SLSN-I sample as this is the most appropriate representation of the corrected peak magnitude root-mean-square that will be available in the future. 

For our DES SN Ia mock Hubble diagram we replicated the approach of \citet{sco16}, who assumed a redshift dependent $\sigma_{\rm err}$, equal to the values published in \citet{Bern12} and reported in Fig. 2 of \citet{sco16}. For DES SNe Ia only, we also included $\sigma_{\rm sys}=0.1$ magnitude to reproduce the possible overall effects of systematic uncertainties \citep[see][for further details]{sco16}.

One systematic uncertainty we must consider is the relative photometric calibration between the local SNe Ia and the more distant SLSN population. Similar to \citet{sco16}, we therefore allowed for an unknown offset between the two samples by including a free parameter $\xi$ in each of the two likelihoods. Therefore, the Hubble residual for the generic $i^{\rm th}$ SN is
\begin{equation}
\Delta \mu_i=\mu_{{\rm obs},i}-\mu_{{\rm cos},i}+\xi,
\end{equation}
where $\mu_{{\rm obs},i}$ is the simulated distance modulus and $\mu_{{\rm cos},i}$ is the theoretical distance modulus using our assumed cosmology. 

We then numerically marginalize over this calibration parameter (by doing so, we also re-absorb any difference in $H_0$ with respect to its fiducial value). This approach is not ideal as it treats a possible systematic uncertainty as an additional statistical noise, but given we are still unclear about the accuracy of any cross-calibration of these samples, it is difficult to model otherwise.

\begin{table*}
\caption{\label{table:cos_constraints} Our one-sigma predicted cosmological constraints for one parameter ($\epsilon = \Delta p/p_{\rm fid}$) for various combinations of likely {\it Euclid} and DES samples and priors (see text for details).}
\centering
\begin{tabular}{lccc}
\hline\hline
Samples & $\Delta\Omega_{\rm m}\left(\epsilon \right)$ & $\Delta w_0 \left(\epsilon\right)$ & $\Delta w_{\rm a}$\\
\hline
& DES SN Ia &&\\
\hline
DES & 0.010 (3\%)& - & - \\
DES & 0.039 (13\%)& 0.119 (12\%)& - \\
DES + P[$\Omega_{\rm m}$] & 0.015 (5\%)& 0.090 (9\%) & 0.47 \\
DES + P[$w_0$] & 0.087 (29\%)& 0.113 (11\%) & 1.03 \\
\hline
& Silver sample (optimistic) &&\\
\hline
DES + {\it Euclid} + low-z & 0.010 (3\%)& - & - \\
DES + {\it Euclid} + low-z & 0.023 (8\%)& 0.085 (9\%)& - \\
DES + {\it Euclid} + low-z + P[$\Omega_{\rm m}$] & 0.014 (5\%)& 0.085 (9\%) & 0.44 \\
DES + {\it Euclid} + low-z + P[$w_0$] & 0.053 (18\%)& 0.083 (8\%) & 0.91 \\
\hline
\end{tabular}
\tablefoot{We do not quote the normalization parameters ($\xi_{\rm SLSN}$ and $\xi_{\rm SNIa}$) as they are irrelevant for the scope of this paper. Priors on the cosmological parameters are given by `P[$\Omega_{\rm m}$]` and `P[$w_0$]' and assumed to be Gaussian of width $\sigma_{\Omega_{\rm m}}=0.015$ ($\sigma_{w_0}=0.25$) on $\Omega_{\rm m}$ ($w_0$) respectively.}
\end{table*}

\subsection{Possible cosmological constraints}
We report the cosmological results for our (optimistic) SLSN-I sample in Table~\ref{table:cos_constraints}. We quote the value of the 1$\sigma$ uncertainties for the free parameters in our fitting ($\Delta p$ in the table, for a generic parameter $p$) which are computed by fitting a Gaussian distribution from the one-dimensional
posterior distributions. We do not quote the best fit values as they are all consistent with our fiducial cosmology within 2$\sigma$. In the same table, we also quote the relative uncertainties $\epsilon=\Delta p/p_{\rm fid}$, for a generic parameter $p$ with fiducial value $p_{\rm fid}$. 
Parameters with a dash symbol (`-') in Table 3 are considered constant within that fit, and fixed to their fiducial values. 

In Table~\ref{table:cos_constraints}, we present results for both a flat $\Lambda$CDM model (assuming $w=-1$) as well as exploring non-zero time derivative of the dark energy equation-of-state parameter, namely $w(a) = w_0 + w_a (1-a)$ \citep{chevallier2001}, which has traditionally been used to quantify possible evolving DE models \citep[e.g. see][]{sola2016,zhao2017}. Such work should show the importance of obtaining high-redshift distance measurements like those discussed here.  

Due to the strong degeneracy between these two dark energy parameters, we therefore include Gaussian priors on $\Omega_{\rm m}$ and $w_0$ of width 0.015 and 0.25, consistent with the current uncertainties found by {\it Planck}. The use of these priors is indicated in Table~\ref{table:cos_constraints} with P[$\Omega_{\rm m}$] and P[$w_0$].

In Table~\ref{table:cos_constraints}, we only show our results in combination with the expected DES SN Ia mock sample as the {\it Euclid} SLSN-I results on their own are not competitive (due to their relatively small numbers and intrinsic scatter presently assumed). For example, in the case of flat $\Lambda$CDM, the {\it Euclid} SLSNe-I alone (plus the low-z SLSN-I sample) would constrain $\Omega_{\rm m}$ to an accuracy of $\sim15\%$. This is not competitive with existing SN-only constraints \citep[e.g.][] {betoule2014}, nor are the results in Table~\ref{table:cos_constraints} when combined with DES for example 3\% uncertainty on $\Omega_{\rm m}$. Again, this is not surprising given the assumed fiducial model, the size of the SLSN-I sample and the assumed uncertainties.  

The cosmological constraints get more interesting when we allow $w$ to vary as the importance of the high-redshift SNe come more pronounced. In Table \ref{table:cos_constraints}, we see that it is possible to improve on the uncertainty of $\Delta w_a$ by adding our 300 SLSNe-I to the existing DES sample (and possible priors from other observations). For example, our overall constraints on the flat $w_0w_a$CDM model (bottom line of Table~\ref{table:cos_constraints}) are better than the best SN-only constraints available in the literature today \citep[e.g. Table 15 of][]{betoule2014}, which show an uncertainty of order one for $w_a$ using {\it Planck}+JLA (and assuming similar weak priors).

We also investigated the possible systematic uncertainty caused by changes in the absolute magnitude of SLSNe-I with redshift due to uncertainties in the $k-$corrections. We repeated the approach in \citet{sco16} of splitting the {\it Euclid} SLSN-I sample into two sub-samples with different normalisation parameters (e.g. different absolute magnitudes). We chose to split the sample at $z=2.5$ as this corresponds to the redshift where the broad UV spectral features in the SLSN-I spectrum shift from the {\it Euclid} $VIS$ band into the IR bands (see Fig.~\ref{fig:sp}). We then analyzed these two sub-samples together with the low-redshift SLSNe-I and the DES SN Ia mock sample, which also has  a free normalization parameter, giving a total of three nuisance parameters in our fitting. We find that both the uncertainties on $\Omega_{\rm m}$ and $w$ increase to 10\% compared to the values given in Table~\ref{table:cos_constraints}. 

The cosmological results in Table~\ref{table:cos_constraints} could improve in several ways. First, our analysis assumes $\sigma_{\rm err}=0.26$ for the dispersion in peak magnitude for our {\it Euclid} SLSNe-I. This is the value obtained by \citet{is14} based on only 14 SLSNe-I available in the literature at the time. If SLSNe-I are standardizable candles, we would expect their standardization to improve in the coming years with on-going surveys and higher quality data on the individual events. In fact, the {\it Euclid} SLSN-I sample should provide an important data set for re-visiting the standardization of these events, and one may wish to include the standardization parameters in the cosmological fitting as presently performed for SN Ia cosmology.

Secondly, we have modeled several possible systematic uncertainties (e.g. lensing, calibration) as additional statistical noise. If these uncertainties could be measured, and corrected for, then we would expect the cosmological constraints to again improve compared to those presented in Table~\ref{table:cos_constraints}. 

\section{Discussion}\label{sec:dis}

We present predictions for the rate of SLSN-I detected by the {\it Euclid} mission. In Fig.~\ref{fig:rates}, we present the expected number of SLSNe-I in the EDS, as a function of redshift, over the nominal five year mission. It is worth stressing that we predict to find a couple of hundred new SLSNe-I to $z\sim3.5$ which will revolutionize our understanding of these enigmatic objects, while providing a new window on the distant Universe. This is possible because of the unique combination of instrumentation available on the {\it Euclid} satellite (a wide-field optical and NIR imager) as well as the EDS observing strategy, for example the continuous monitoring of the same field, which minimizes temporal edge effects that may affect ground-based searches. As demonstrated in the large uncertainty on our predictions, these {\it Euclid} data will immediately provide a precise determination of the SLSN-I rate (with redshift), thus helping constraint the star-formation history of the Universe at these early epochs. There is no other experiment prior to {\it Euclid} that will provide such information on high-redshift SLSNe-I, and therefore it is important to use these unique data to the best of our ability (especially if observed contemporaneously with other facilities like LSST). 

In addition to improving our understanding of the astrophysics of these objects, the {\it Euclid} SLSN-I sample provides additional cosmological constraints as discussed in Sect.~\ref{sec:cosmo}. These constraints will be complementary to those planned from {\it Euclid} weak gravitational lensing, galaxy clustering and SNe Ia (e.g. DESIRE), as well as probing to higher redshift than SLSN-I samples from LSST. For example, in \citet{sco16}, we presented cosmological constraints from an idealised sample of LSST SLSNe-I, but the unavailability of deep NIR imaging over the LSST area will limit the detection of SLSNe beyond $z\sim2$. These {\it Euclid} SLSNe-I will therefore be unique in allowing us to extend the overall cosmological constraints to $z\sim3.5$ thus improving our constraints on possible dynamical DE models \citep{sola2016,zhao2017}. Any additional high-redshift measurements of the expansion history of the Universe are welcome, especially if they come for free from data already planned. 

The results of this paper depend on obtaining spectra for as many {\it Euclid} SLSNe-I as possible (see Sect.~\ref{sec:astro}). This implies the need for an effective (in terms of purity and completeness) method for identifying as many as `true' SLNSe-I from other types of transients. In the case of objects not matching the conservative selection criteria used to define the golden and silver samples, other methods may be effective such as that presented by \citet{disanto2016}. In this case, light-curves are first compressed to a reduced, but extensive, set of statistical features and then classified using the {\tt MLPQNA} method \citep{bres14}. While this approach has never been applied to the classification of SLSNe-I, it has achieved  96\% completeness and 85\% purity in classifying SNe Ia in the Catalina Real-Time Transient Survey. 

Finally, we raise the possibility of measuring the gravitational lensing of these high-redshift SLSNe-I as we did include it as a likely `noise' term in our analysis above. While the probability of witnessing a strongly lensed SLSN-I \citep{kelly2015} will be small, it may be possible to measure the cross-correlation function between the peak, corrected magnitudes of these distant SLSN-I with the foreground large-scale structure as traced by galaxies \citep{scovacricchi2016_b}, especially as these {\it Euclid} deep fields will become the focus of significant additional observations for example there will likely be overlap with LSST deep drill fields, and possibly WFIRST \citep{wfirst}.

\section{Conclusions}\label{sec:con}
We present an analysis of the possible number of superluminous supernovae detected in the {\it Euclid} Deep  Survey. We show that {\it Euclid} should find $\simeq 140$ high-quality SLSNe-I to $z \sim 3.5$ over a five year period. An extra $\simeq70$ SLSNe-I are possible depending on the quality cuts, while present uncertainties in the rates, luminosity functions, and instrument detection efficiencies may allow many more to be found. These data, especially if also spectroscopically targeted to secure their nature and redshift, will revolutionize the study of SLSNe-I, increasing present samples of high-redshift ($z>2$) SLSN-I by two orders of magnitude. 
 
We stress the importance of these {\it Euclid} SLSNe-I for the study of supernova astrophysics and the star-formation history of the Universe. Such investigations will be enhanced by follow-up observations by the next generation of large space and ground--based telescopes (E-ELT, LSST, JWST) and provide excellent targets for these observatories. {\it Euclid} will also provide low-resolution NIR grism spectroscopy for some low-redshift SLSNe-I.

We also investigated the possibility of constraining cosmology using these SLSNe-I, when combined with a low-redshift sample of 50 SLSNe-I (from the literature), and the expected cosmological results from DES. In the case of a flat $w_0 w_a$CDM model, our analysis suggests we could obtain an uncertainty of $\Delta w_a\sim0.9$ which is an improvement on DES alone result, and the present constraints on this parameterization. Any additional measurements of the high-redshift expansion history of the Universe are invaluable as present baryonic acoustic oscillations  observations suggest a possible tension with the standard $\Lambda$CDM model \citep{zhao2017}, either indicating unrecognized systematic uncertainties or dynamical dark energy. 

We finish by noting that these {\it Euclid} SLSNe-I come `for free' as we have just assumed the latest survey design of the EDS. It is therefore important to prepare the {\it Euclid} analysis software pipelines to detect such transients as they will be present in the data. This is a major motivation for this paper, that is, to highlight the urgent need to prepare for such long-lived transients in the {\it Euclid} data-stream and be ready to detect them in `real-time' \citep[within days hopefully, but see][about SLSN classification in real time]{in17b} to trigger follow-up observations, for example using JWST which also has a finite lifetime. 

\begin{acknowledgements}
We thank the internal EC referees (P. Nugent and J. Brichmann) as well as the many comments from our EC colleagues and friends. 
CI thanks Chris Frohmaier and Szymon Prajs for useful discussions about supernova rates. CI and RCN thank Mark Cropper for helpful information about the $VIS$ instrument. CI thanks the organisers and participants of the Munich Institute for Astro- and Particle Physics (MIAPP) workshop `Superluminous supernovae in the next decade' for stimulating discussions and the provided online material.
The {\it Euclid} Consortium acknowledges the European Space Agency and the support of a number of agencies and institutes that have supported the development of {\it Euclid}. A detailed
  complete list is available on the {\it Euclid} web site 
(\texttt{http://www.euclid-ec.org}). In particular the Agenzia Spaziale
  Italiana, the Centre National dEtudes Spatiales, the Deutsches
  Zentrum f\"ur Luft- and Raumfahrt, the Danish Space Research
  Institute, the Funda\c{c}\~{a}o para a Ci\^{e}nca e a Tecnologia,
  the Ministerio de Economia y Competitividad, the National
  Aeronautics and Space Administration, the Netherlandse
  Onderzoekschool Voor Astronomie, the Norvegian Space Center, the
  Romanian Space Agency, the State Secretariat for Education, Research
  and Innovation (SERI) at the Swiss Space Office (SSO), the United
  Kingdom Space Agency, and the University of Helsinki.\xspace
RCN acknowledges partial support from the UK Space Agency. DS acknowledges the Faculty of Technology of the University of Portsmouth for support during his PhD studies. CI and SJS acknowledge funding from the European Research Council under the European Union's Seventh Framework Programme (FP7/2007-2013)/ERC Grant agreement n$^{\rm o}$ [291222]. CI and MS acknowledge support from EU/FP7-ERC grant no [615929]. EC acknowledge financial contribution from the agreement ASI/INAF/I/023/12/0. The work by KJ and others at MPIA on NISP was supported by the Deutsches Zentrum f\"ur Luft- und Raumfahrt e.V. (DLR) under grant
50QE1202. MB and SC acknowledge financial contribution from the agreement ASI/INAF I/023/12/1. RT acknowledges funding from the Spanish Ministerio de Econom\'ia y Competitividad under the grant ESP2015-69020-C2-2-R. IT acknowledges support from Funda\c{c}\~ao  para  a  Ci\^encia e a Tecnologia (FCT) through the research grant UID/FIS/04434/2013 and IF/01518/2014. JR was  supported by JPL, which is run under a  contract for NASA by Caltech and by NASA ROSES  grant 12-EUCLID12-0004.

\end{acknowledgements}

\end{document}